\documentclass{PoS}

\usepackage{amsmath}
\usepackage{amssymb}
\usepackage{array}
\usepackage{calc}
\usepackage{longtable}
\usepackage{multirow,booktabs}
\usepackage{color}
\usepackage{relsize}
\usepackage{pstricks}
\usepackage{graphicx}
\usepackage{xspace}
\usepackage{units}

\newcommand{\MEPSatNLO}{M\protect\scalebox{0.8}{E}P\protect\scalebox{0.8}{S}@N\protect\scalebox{0.8}{LO}\xspace}

\newcommand{\NLO}{N\protect\scalebox{0.8}{LO}\xspace}

\newcommand{\Sherpa}{S\protect\scalebox{0.8}{HERPA}\xspace}

\title{Heavy quark mass effects in associated production}

\ShortTitle{Heavy quark mass effects in associated production}

\author{\speaker{Davide Napoletano}\\
        IPhT, CEA Saclay, CNRS UMR 3681, F-91191, Gif-Sur-Yvette, France\\
        E-mail: \email{davide.napoletano@ipht.fr}}

\abstract{Processes involving heavy quarks in associated production, 
    are usually described in two factorisation schemes: one in which the heavy quark
    is treated as an infinitely massive object, decoupled from QCD evolution,
    and one in which it is treated on the same footing of a light quark.
    These two approaches only differ by the inclusion of mass suppressed terms
    or in the resummation of a certain class
    of logarithms. In view of recent results,
    in this talk, we present recent developments that extend a phenomenological 
    scheme, that tries to incorporate the best of both worlds, to \NLO in Monte
    Carlo event generation.}

\FullConference{XXVI International Workshop on Deep-Inelastic Scattering and Related Subjects (DIS2018)\\
		16-20 April 2018\\
		Kobe, Japan}

\begin{document}
The production of heavy quarks in association with a vector boson, or a Higgs
boson, has re-attracted interest from the theory community in recent years~\cite{Maltoni:2012pa,
    Wiesemann:2014ioa,Forte:2015hba,Bonvini:2015pxa,Krauss:2016orf,Bonvini:2016fgf,Lim:2016wjo,
    Forte:2016sja,Krauss:2017wmx,Bagnaschi:2018dnh,Forte:2018ovl} for essentially two
reasons. Firstly, they constitute an irreducible background for 
many Higgs processes, both in the Standard Model and beyond, which requires good
theoretical control for calculations of such processes.
The second, and more theoretical reason, comes from the
fact that these processes involve at least two, well separated, scales. As it is well
known, this manifests itself, in QCD processes, through logarithms of ratios of these
scales, which one may or may not want to resum.

When, in particular, $b$ quarks are excited from the proton, and thus
appear in the initial state, two factorisation schemes are usually employed,
the 4-flavour and the 5-flavour schemes.
In the 4-flavour scheme, the $b$ quark is considered as an infinitely massive
object decoupled from QCD evolution. In this approach, then, it can
only be produced in the final state, through a $g\rightarrow b\bar{b}$ splitting
if the gluon has enough energy. Fixed-order logarithms of any energy-like invariant
over $m_b$ can appear.
In the 5-flavour scheme $b$ quarks have zero mass, and thus participate in the 
QCD evolution, like all other light quarks. $b$-mass effects, appear only as
threshold effects, and logarithms of $\mu_F/m_b$ are resummed to all order.
\begin{figure}[htb]
  \centering{%
    \includegraphics[scale=0.4,angle=270]{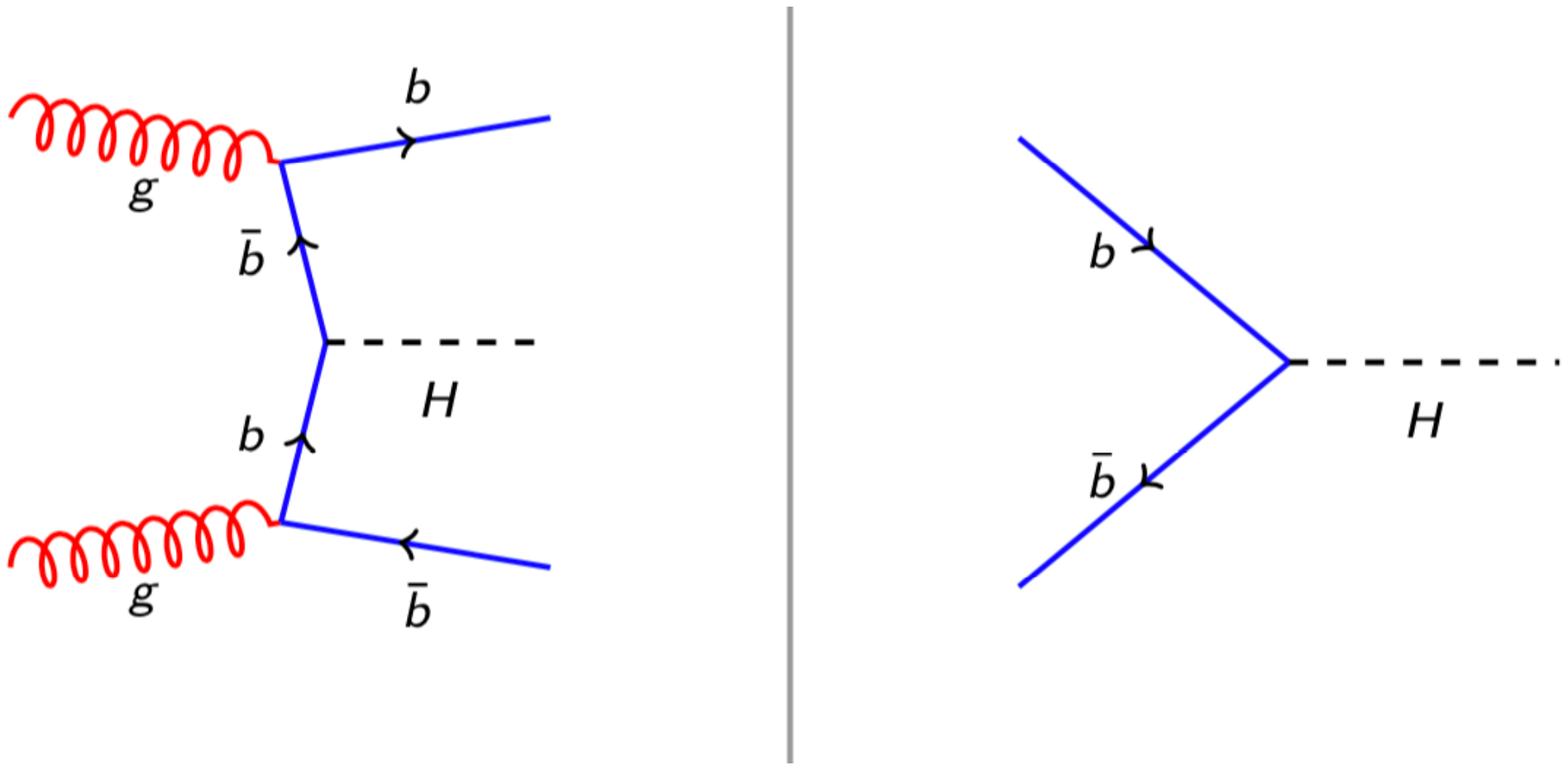}}
  \caption{Sketched comparison of the 4F and the 5F scheme.}
  \label{Fig:4fvs5fscheme}
\end{figure}

In inclusive enough observables, and easy enough processes, like in the calculation of the total inclusive
cross section in $b\bar{b}\rightarrow H/Z$, setting $\mu_F$ in the 5 flavour scheme to the value
of the hard scale in the four flavour scheme, one can easily see the correspondence
between the two schemes (Fig.~\ref{Fig:4fvs5fscheme}) and construct a matching scheme to combine the two
and obtain a result which includes both fixed-order mass effects, and the 
resummation of large $\mu_F/m_b$ logarithms~\cite{Forte:2015hba,Bonvini:2015pxa,Bonvini:2016fgf,
    Forte:2016sja,Forte:2018ovl}.

\begin{figure}[!htb]
  \begin{center}
      \includegraphics[width=0.45\textwidth]{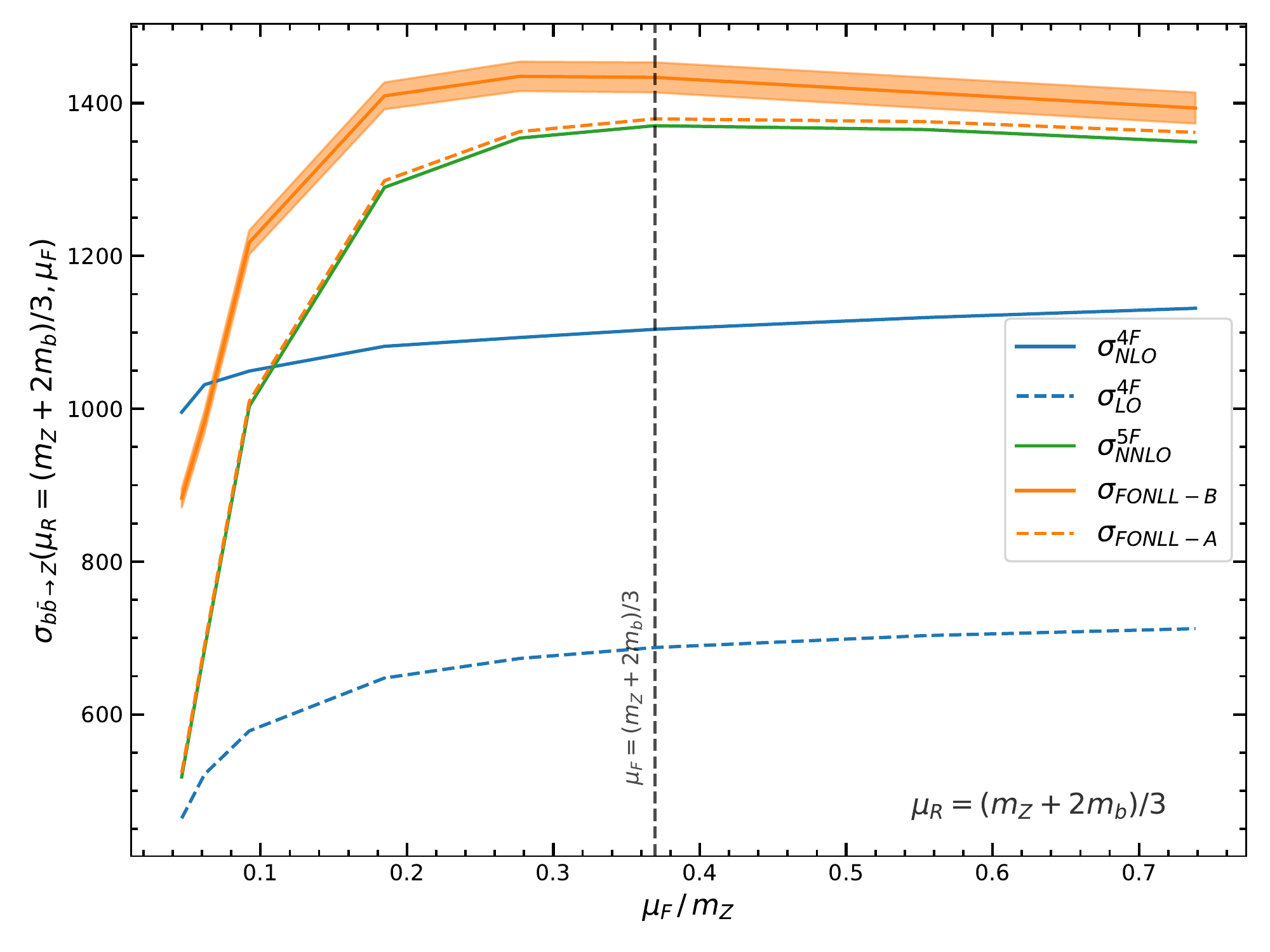}
      \includegraphics[width=0.45\textwidth]{muF_var.pdf}\\
      \vspace{-1.5cm}
      \includegraphics[width=0.5\textwidth]{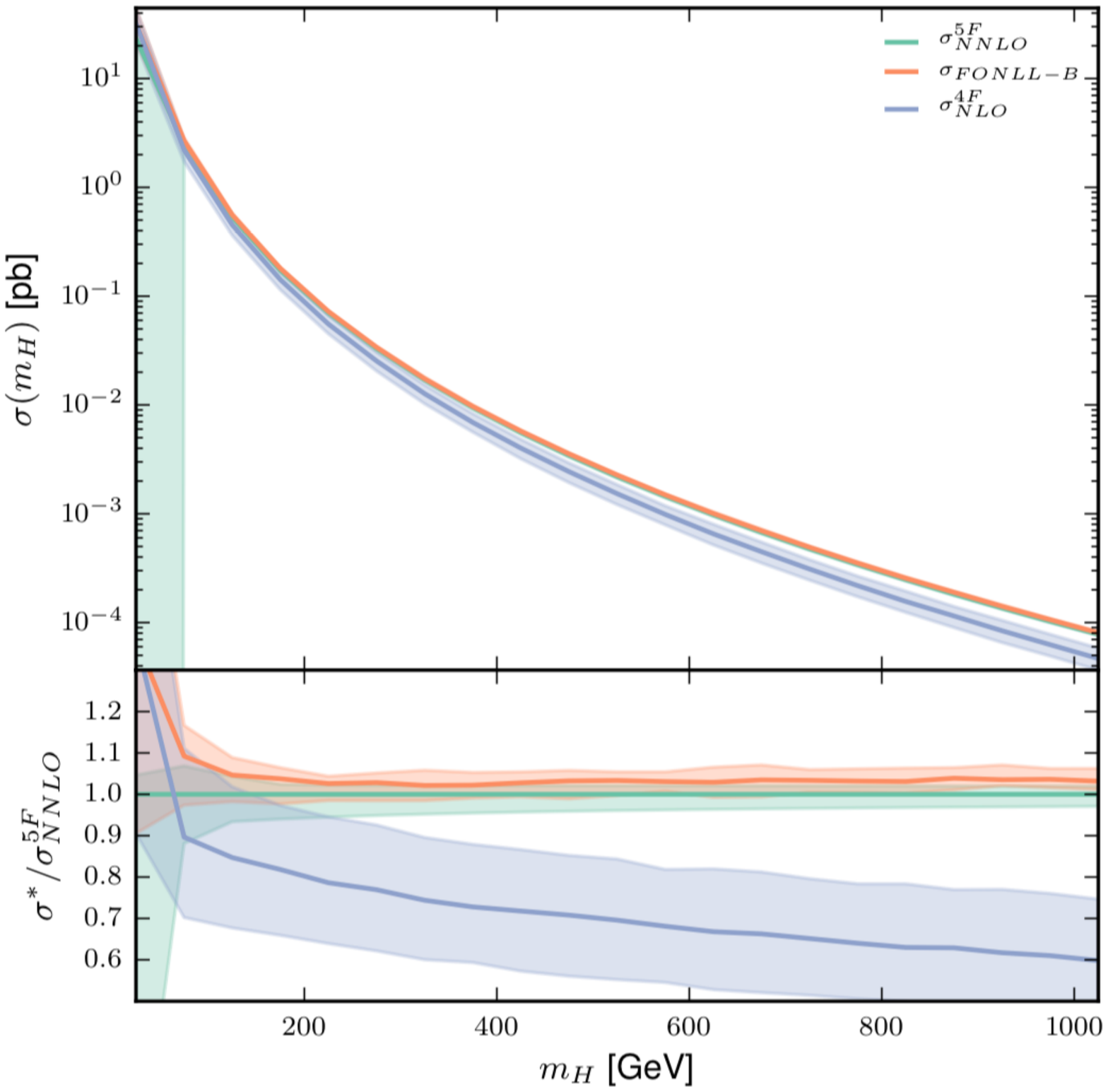}
      \vspace{-1.5cm}
    \caption{The FONLL matching scheme for the inclusive cross section
    of $b\bar{b} \rightarrow Z$ (Top row) as a function of $\mu_{R,F}$ and
    of $b\bar{b} \rightarrow H$ (Bottom row) as a function of $m_H$. Matched
    results are compared to the standard 4F and 5F schemes.}
  \label{fig:bbzfonll}
  \end{center}
\end{figure}

These matched results can give insights on what are the main contributions
that make the two factorisation schemes different: mass or resummation effects.
Looking, for example, at the top row of Fig.~\ref{fig:bbzfonll}, one
can see that the matched result, both with leading or next-to-leading
fixed order accuracy, lies very close to the nominal five flavour scheme,
thus concluding that mass effects play a little role, with the main difference
between the two schemes, coming from the resummation of large logs. 

The conclusion just shown seems to generally hold even in differential distributions
for more exclusive observables~\cite{Krauss:2016orf}. Nevertheless, 
there are some regions of phase-space where the inclusion
of fixed-order mass effects is more important than resummation. However, 
performing an analytical matching in these more exclusive cases is typically
just not doable. For this reasons, we propose to extend a old idea, used mostly
in DIS processes, to hadron-hadron collisions at \NLO accuracy: retaining mass
effects in the initial state for heavy quarks. We name this scheme, the five flavour massive
scheme, or 5MS in short~\cite{Krauss:2017wmx}.

The main idea of this scheme, is to allow $b$ (or any other heavy) quarks in the initial state 
as in the 5F scheme, yet retaining full mass dependence at the matrix element and phase-space
level. This is done at \NLO by extending Catani-Seymour~\cite{Catani:1996vz,Catani:2002hc,Dittmaier:1999mb} 
subtraction in the \Sherpa Monte
Carlo event generator, to allow for massive initial state~\cite{Krauss:2017wmx,Gleisberg:2008ta}.

\begin{figure}[htb]
  \centering{%
    \includegraphics[width=0.6\textwidth]{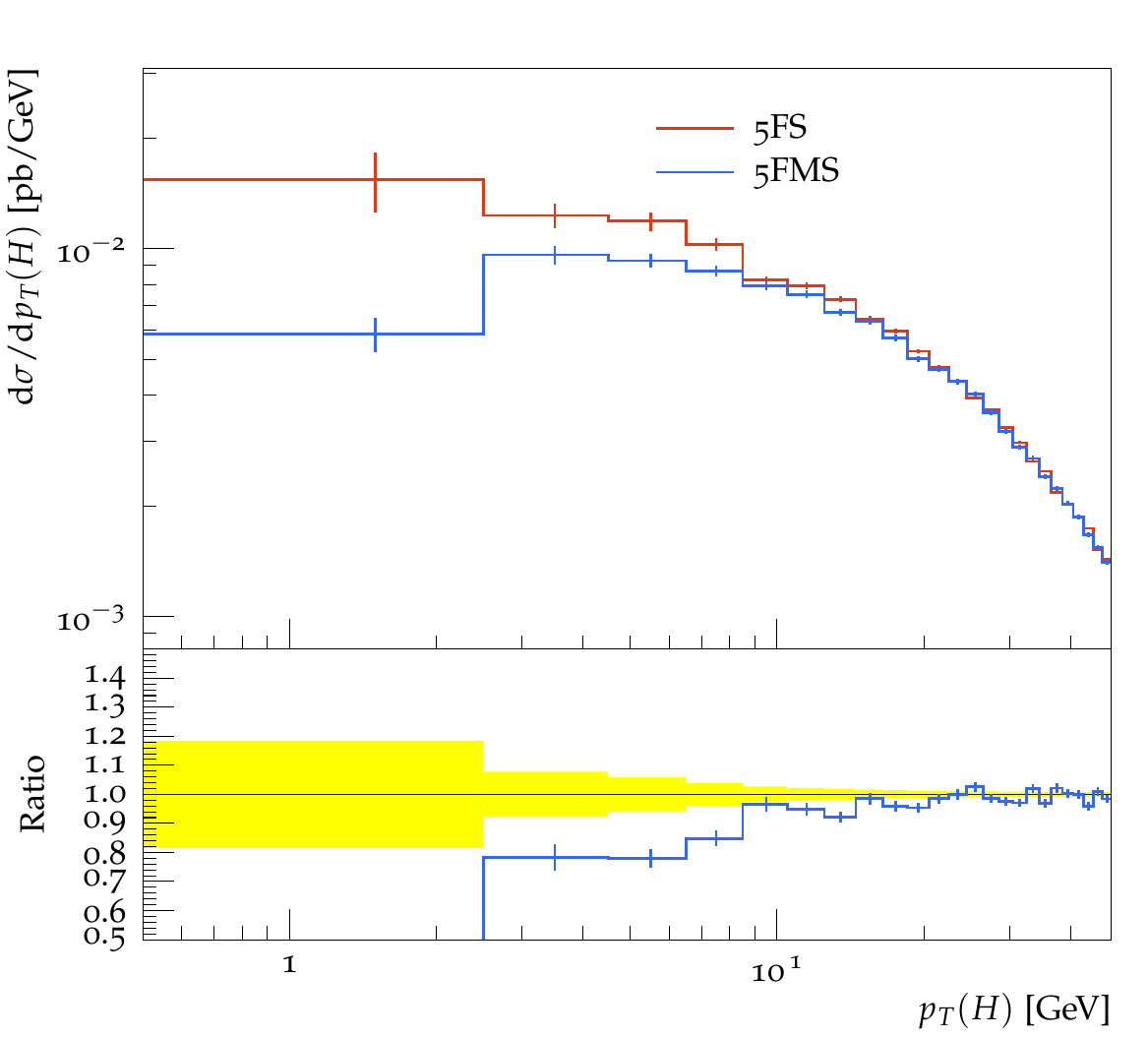}}
  \caption{Comparison of the 5F and the 5FM scheme in $b\bar{b} \rightarrow H$ at \NLO fixed order accuracy.}
  \label{Fig:5F5FMS_fo}
\end{figure}

\begin{figure}[!htb]
  \begin{center}
    \begin{tabular}{cc}
      \includegraphics[width=0.4\textwidth]{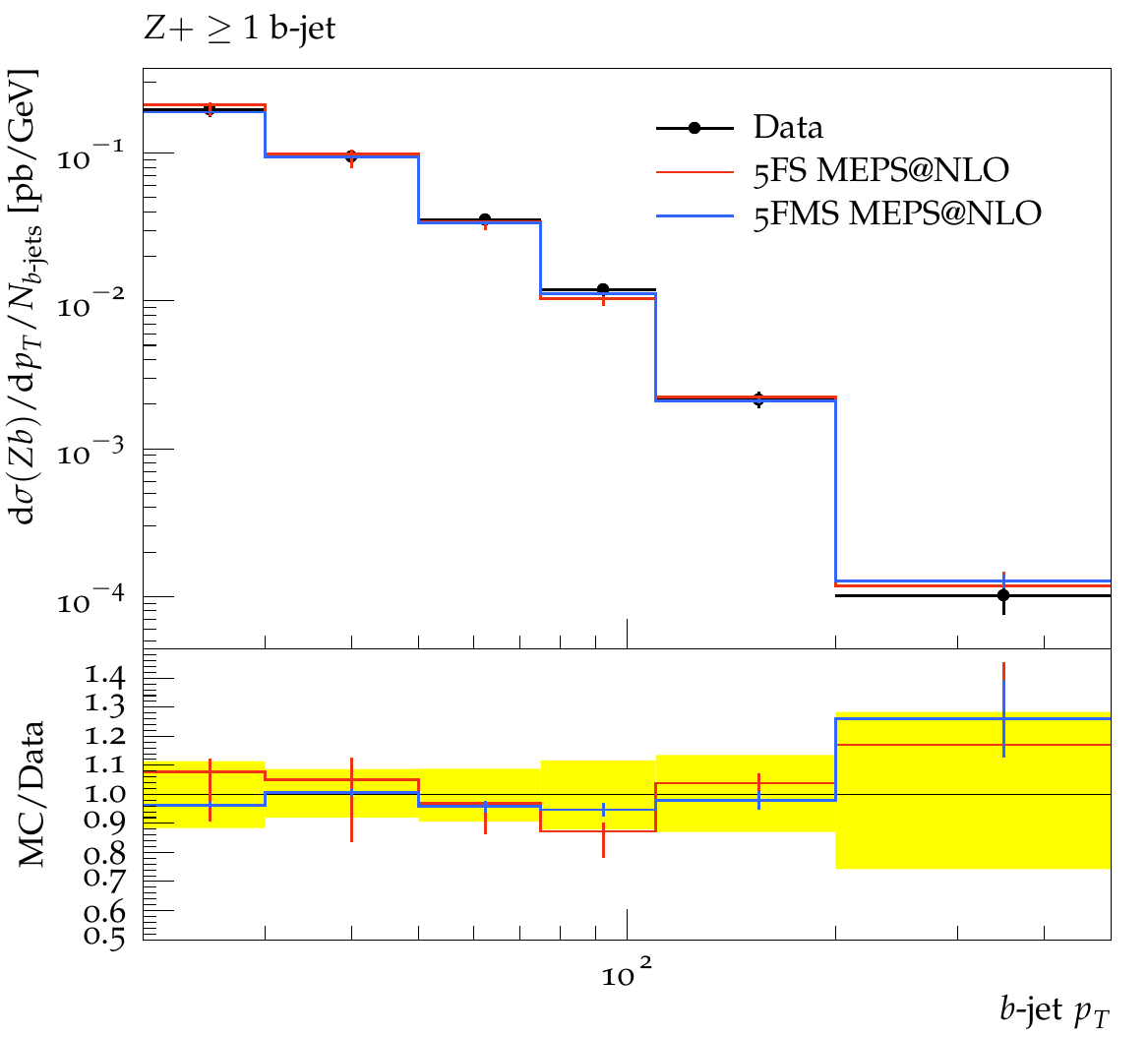}&
      \includegraphics[width=0.4\textwidth]{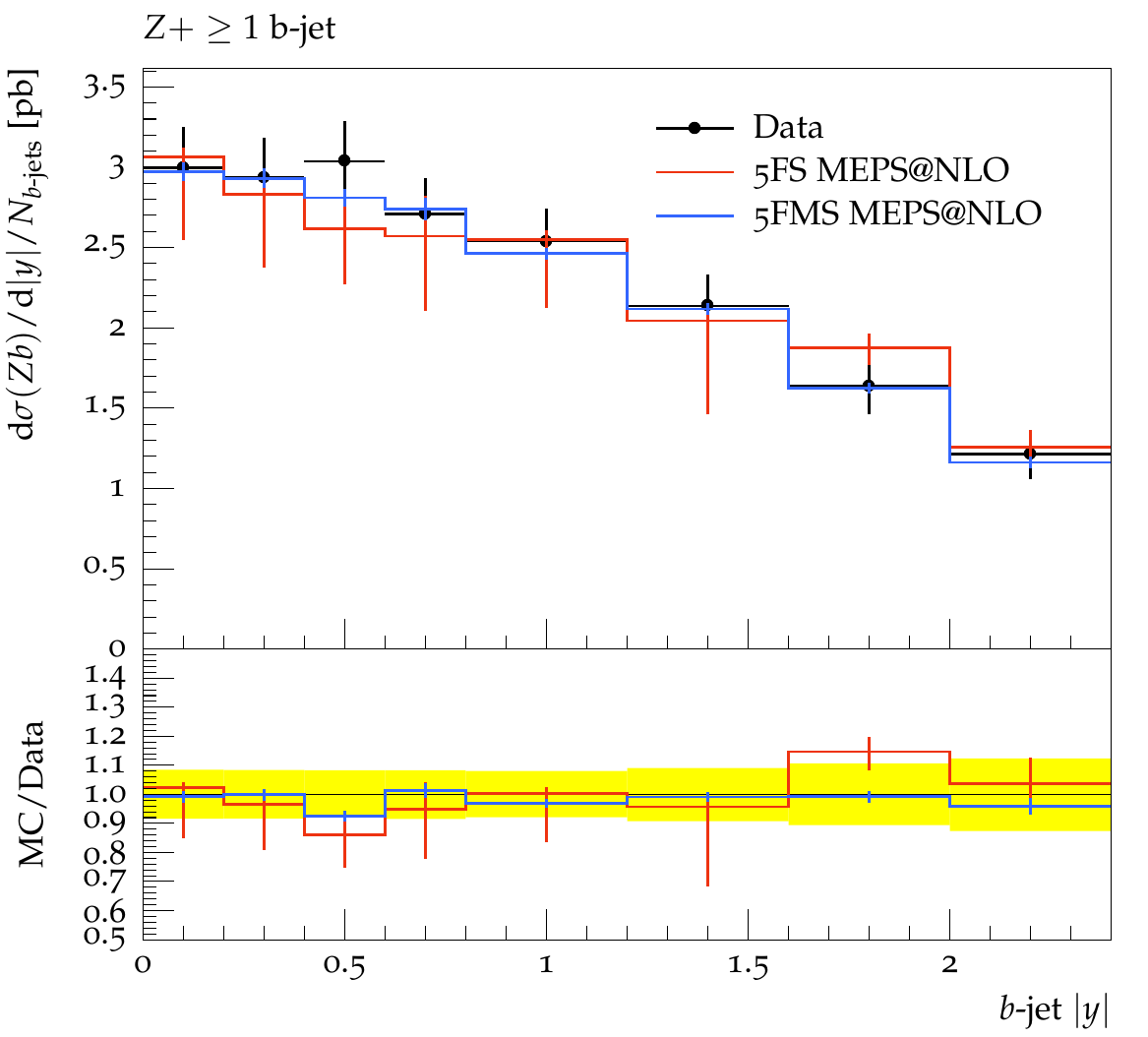}\\
      \includegraphics[width=0.4\textwidth]{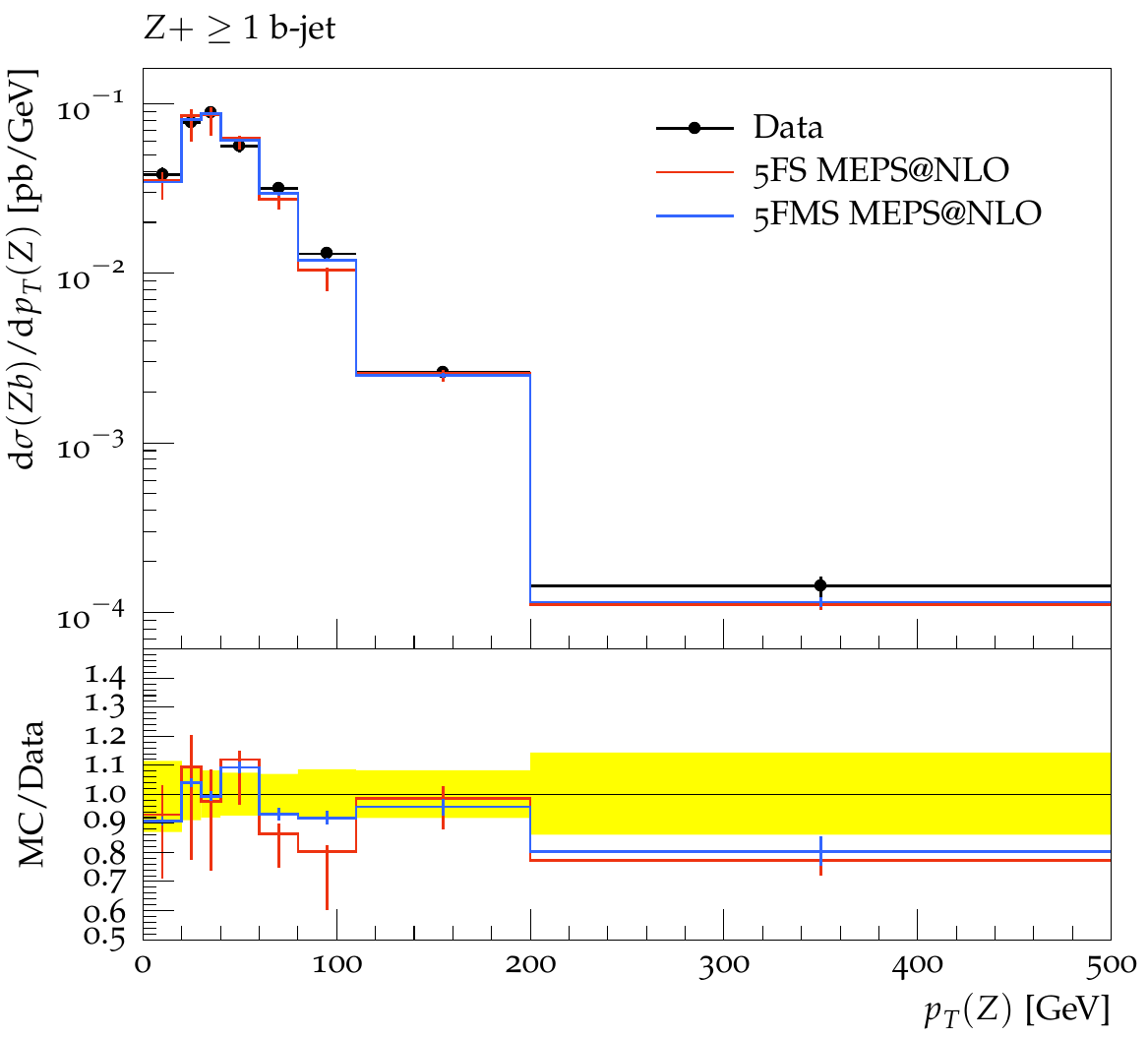}&
      \includegraphics[width=0.4\textwidth]{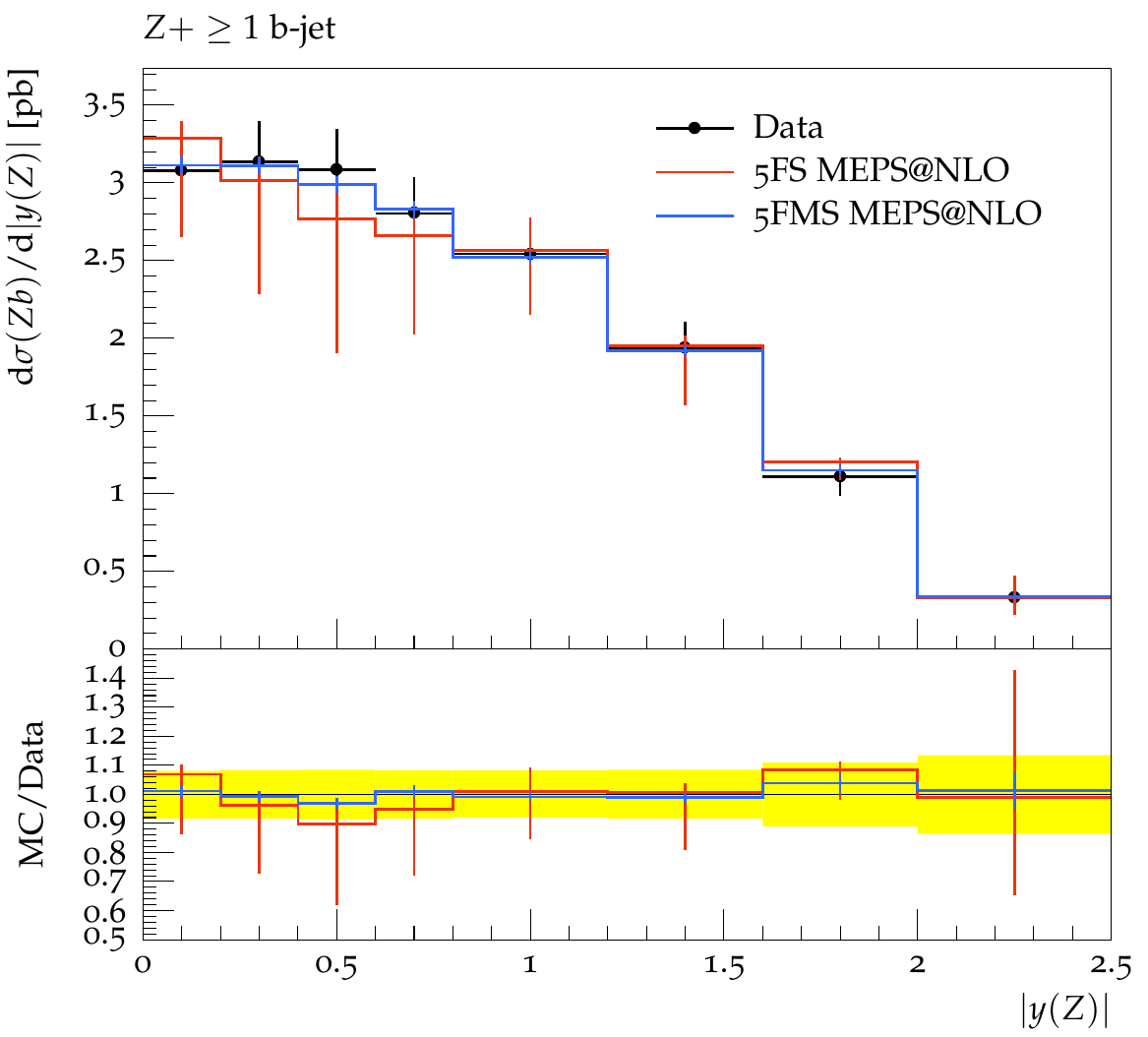}     
    \end{tabular}
    \caption{We show prediction obtained
      in the 5FS, massless, at \MEPSatNLO accuracy, with up to 2 jets at NLO
      plus up to three jets at leading order. The 5FMS prediction on the other hand
      includes only the 0 jet contribution at NLO, while the 1,2 and 3 jets
      contributions are merged with LO accuracy, data from~\cite{Aad:2014dvb}}
  \label{fig:mcatnlo}
  \end{center}
\end{figure}

To see how the inclusion of such mass effects affects differential predictions, we
studied for $b\bar{b} \rightarrow H$ at \NLO fixed order accuracy and
$b\bar{b} \rightarrow Z$ in a \MEPSatNLO~\cite{Hoeche:2010kg} simulation. Results are reported
in Figs.~\ref{Fig:5F5FMS_fo} and \ref{fig:mcatnlo} respectively. 

In Fig.~\ref{Fig:5F5FMS_fo} it can be seen how mass effects have an impact,
and also quite a large one, only in the very low $p_T$ region.
Once $p_T$ becomes of the order of about twice the bottom mass, these
effects become negligible and the massless and massive schemes converge to
each other, as expected.

Moving to Fig.~\ref{fig:mcatnlo}. Here we generate a merged~\cite{Catani:2001cc} sample of 
$Z$ plus 0,1,2 and 3 jets respectively. Further we look at events with
at least one $b$ jet. This means that, on top of the bottom initiated contributions,
we also have the light quark ones. The latter are typically much larger than the former, 
as their parton distribution function is much larger. In addition, because of experimental statistics,
very low $p_T$ bins are integrated into larger bins than the ones shown for the
fixed order case. For this reasons in this process we do not see any significant
effect coming from the inclusion of mass effects.

Further details and studies on this proposed scheme can be found in~\cite{Krauss:2017wmx}

\bibliographystyle{unsrt}
\bibliography{bbz_fonll}

\end{document}